\documentclass[a4paper,11pt]{article}
\usepackage[utf8]{inputenc}
\usepackage[T1]{fontenc}
\usepackage{a4wide,amsmath,amssymb,epsfig}
\usepackage{graphicx}
\usepackage{authblk}
\begin{document}
\title{On certain analytical methods in finding integrable systems and their interconnections}
\author{R. Mohanasubha$^1$, V. K. Chandrasekar$^2$, M. Senthilvelan$^1$\footnote{E-mail: senv0000@gmail.com} \\and M. Lakshmanan$^1$}
\affil{1. Centre for Nonlinear Dynamics, School of Physics,
Bharathidasan University, Tiruchirappalli - 620 024, India\\
2. Centre for Nonlinear Science and Engineering, School of Electrical and Electronics Engineering, SASTRA University, Thanjavur - 613 401, India}
\maketitle
\begin{abstract}
In this paper, to begin with, we review six different analytical methods which are widely used to derive symmetries, integrating factors, multipliers, Darboux polynomials and integrals of second order nonlinear ordinary differential equations. We illustrate the algorithm behind each method by considering a nonlinear oscillator equation as an example. In the second part of this paper, we examine the interconnections between these methods. We establish a road map between extended Prelle-Singer procedure with all other methods cited above and thereby demonstrate the interplay between Lie point symmetries, $\lambda$-symmetries, adjoint symmetries, null forms, integrating factors, Darboux polynomials, and Jacobi multipliers of second order integrable ODEs. The interconnections are illustrated with the same example finally. 
\end{abstract}

\section{Introduction}
During the past few years there has been a great deal of interest in finding the interrelation between symmetries, integrating factors, integrals of motion and integrability of the associated dynamical systems, especially for finite degrees of freedom systems \cite{bab}. Confining our attention on dynamical systems described by second order linear/nonlinear ordinary differential equations, the integrability is guaranteed either by providing a time independent integral (which can be interpreted as the Hamiltonian) or two time dependent integrals \cite{mlsr}. The required integrals can be constructed with the help of symmetries (it may be Lie point symmetries, adjoint symmetries, $\lambda$-symmetries, contact symmetries or nonlocal symmetries), integrating factors and multipliers. To investigate these quantities, that is symmetries, integrating factors and integrals, several powerful methods have been developed in the literature over a long period \cite{olv,mur,blum}. The methods that are widely used to explore these quantities are the following (chronological order): (i) Jacobi last multiplier, (ii) Darboux polynomials, (iii) Lie point symmetries, (iv) adjoint symmetries, (v) $\lambda$-symmetries and (vi) Prelle-Singer method. We note here that we have listed out only the popular methods in the literature and left the others which are applicable only for specific equations.  

Since we are going to concentrate only on these methods, in the following, we recall briefly the development of each method. The Jacobi last multiplier method was introduced by Jacobi in the year 1844 \cite{jaco1,jaco2}. The method lay dormant for several decades. Only recently Nucci and her collaborators have demonstrated the applicability of this method in exploring non-standard Lagrangians associated with certain second order nonlinear ODEs \cite{nuc1,nuc2,nuc3}. One of the important applications of multipliers is that one can determine the integrals associated with the given equation by evaluating their ratios. 

Darboux polynomials method was introduced by Darboux in the year 1878 \cite{dar1}. There exist a lot of different names in the literature for Darboux polynomials, for example we can find: special integrals, eigenpolynomials, algebraic invariant curves, particular algebraic solutions or special polynomials. The method also provides a strategy to find first integrals. If we have enough Darboux polynomials then there exists a rational first integral. Darboux showed that if we have $\frac{d(d+1)} {2}+2$ Darboux polynomials, where $d$ is the order of the given equation, then there exists a rational first integral which can be expressed in terms of these polynomials \cite{chav1,chri,lib1,lib4}.

Lie symmetry analysis is one of the powerful methods to investigate the group invariance properties of the given dynamical system, and establishing the integrability of the given differential equation is just a by-product of this method. The method was developed by Sophus Lie at the end of nineteenth century. In this method one first identifies the symmetry vector fields associated with the differential equation. The symmetry vector fields can then be used to derive integrating factors, conserved quantities, similarity variables and so on \cite{olv}.

Exploring integrating factors and obtaining integrals for the given equation is one of the classical  ideas and there exists several methods in the literature to obtain the integrating factors. One such method is the adjoint symmetry method. This method was developed by Bluman and Anco in the year 1998 \cite{blum}. 

All the nonlinear ODEs do not necessarily admit Lie point symmetries.  Under such a circumstance one may look for  generalized symmetries (other than Lie point symmetries) associated with the given equation.  One such generalized symmetry is the $\lambda$-symmetry.  The $\lambda$-symmetries can be derived by a well defined algorithm which includes Lie point symmetries as a specific subcase and have an associated order reduction procedure which is similar to the classical Lie method of reduction \cite{mur}.  Although $\lambda$-symmetries are not Lie point symmetries, the unique prolongation of vector fields  to the space of variables $(t,x,...,x^n)$ for which Lie reduction method applies is always a $\lambda$-prolongation for some function $\lambda(t,x,\dot x)$. For more details, one may see the works of Muriel and Romero \cite{mur}.

Recently, a powerful method, namely the extended Prelle-Singer method was developed by our group \cite{chan}. The method is an extended and complete picture of the technique given by Duarte et al. for second order ODEs \cite{duar}. The original method, which is applicable to planar ODEs, was proposed by Prelle-Singer in the year 1983 \cite{prelle}. The attractiveness of the Prelle-Singer (PS) method is that the method guarantees that a solution will be found if the given system of first-order ODEs has a solution in terms of elementary functions. The same methodology has also been demonstrated in second and higher order nonlinear ODEs including coupled ones \cite{chan}. 

Even though the aforementioned methods have been introduced at different times in the literature recently attempts have been made to investigate the interconnection between symmetries, multipliers, integrating factors and integrals  as well as the methods that produce them. Upon analyzing the factors the interconnection have been found between (i) $\lambda$-symmetries and Lie point symmetries \cite{mur}, (ii) Jacobi last multiplier with Darboux polynomials \cite{car2}, (iii) $\lambda$-symmetries and null forms \cite{mur} and (iv) Lie point symmetries and Jacobi last multiplier \cite{nuc1}. A natural question which arises here is whether there exists a more encompassing interconnection which relates all these methods/quantities. In this paper we establish a road map between extended Prelle-Singer procedure with all other methods cited above and thereby demonstrate the interplay between Lie point symmetries, $\lambda$-symmetries, adjoint symmetries, null forms, integrating factors, Darboux polynomials, and Jacobi multipliers of second order integrable ODEs. For more details, see our recent work \cite{subha}.

We organize the presentation as follows: In Section 2, we give an overview about these methods with an example. In Section 3, we give the interlinks between these methods. We also demonstrate the interlinks with the same example. Finally, we present our conclusions in Sec.4.
\section{Analytical methods}
In this section we briefly recall six analytical methods which are widely used in the contemporary literature to derive the following quantities, namely (i) integrals, (ii) symmetries, (iii) integrating factors and (iv) multipliers. To make our presentation concise and comparable we consider the same example in all the methods and illustrate the ideas behind each method. 
\subsection{Jacobi last multiplier (JLM)~\cite{jaco1,jaco2}}
Let us consider a second order ODE
\begin{eqnarray}
\ddot{x}=\phi(t,x,\dot{x}),\label{main}
\end{eqnarray}
where $\phi$ is an analytic function of the variables $t$, $x$ and $\dot{x}$. Let us rewrite the second order ODE (\ref{main}) into a system of two first-order ODEs:
\begin{eqnarray}
\frac{dx}{dt}=f(x,y),~~\frac{dy}{dt}=g(x,y).\label{jlmere}
\end{eqnarray}
Dividing the second expression by the first expression of (\ref{jlmere}) and multiplying the resultant equation by an integrating factor $M(x,y)$ we arrive at
\begin{eqnarray}
M(gdx-fdy)=0\Rightarrow dI=0.\label{ijkl}
\end{eqnarray}
Let us assume that Eq.(\ref{jlmere}) has a first integral $I(x,y)$ which is constant on the solutions. In this case, we have
\begin{equation}
dI=\frac{\partial I} {\partial x}dx+\frac{\partial I} {\partial y}dy=0.\label{jlm21}
\end{equation}
Comparing Eqs.(\ref{jlm21}) and (\ref{ijkl}), we find $\frac{\partial I} {\partial x}=Mg$ and $\frac{\partial I} {\partial y}=-Mf$. By imposing the compatibility condition, $\frac{\partial^2 I} {\partial x \partial y}=\frac{\partial^2 I} {\partial y \partial x}$, we get $M_yg+Mg_y=-M_xf-Mf_x$. The later equation can be rewritten as
\begin{eqnarray}
\frac{D[M]} {M}+f_x+g_y=0\Rightarrow D[\log M]+f_x+g_y=0,~~~D=\frac{\partial}{\partial t}+\dot{x}\frac{\partial}{\partial x}+\dot{y}\frac{\partial}{\partial y}.\label{jljkklij}
\end{eqnarray}

Choosing the functions $f$ and $g$ suitably, the determining equation for the Jacobi last multiplier for a second order ODE can be written as
\begin{equation}
D[\log M]+\phi_{\dot{x}}=0.\label{met20}
\end{equation}
Solving Eq.(\ref{met20}) we can obtain the Jacobi last multipliers $M$. Once sufficient number of Jacobi last multipliers are determined the associated first integral can be found from their ratios.

To demonstrate that the ratio of two multipliers defines an integral, let us assume that $M_1$ and $M_2$ are two Jacobi last multipliers. In this case, we have
\begin{equation}
D[\log M_1]+\phi_{\dot{x}}=0,~~~D[\log M_2]+\phi_{\dot{x}}=0.
\end{equation}
Let $I$ be the ratio of two Jacobi last multipliers, that is $I=\frac{M_1} {M_2}$. Evaluating the total derivative we find
\begin{eqnarray}
\frac{dI} {dt}=\frac{d}{dt}\bigg(\frac{M_1}{M_2}\bigg)=0&=&\frac{\dot{M_1}M_2-M_1\dot{M_2}} {M_2^2}=\frac{\dot{M_1}}{M_1}-\frac{\dot{M_2}}{M_2}\nonumber \\
&=&\frac{d} {dt}\log M_1-\frac{d} {dt}\log M_2=\phi_{\dot{x}}-\phi_{\dot{x}}=0.\nonumber
\end{eqnarray}
Thus if we have sufficient number of multipliers we can obtain the necessary integrals.

\subsubsection{Example}
Let us illustrate the above method with the following second order nonlinear ODE, namely
\begin{equation}
\ddot{x}-\frac{3}{2}\frac{\dot{x}^2}{x}+2x^3=0.\label{exam}
\end{equation}
The Jacobi multipliers of Eq.(\ref{exam}) can be determined by integrating the following equation, that is $M_t+\dot{x}M_x+(\frac{3}{2}\frac{\dot{x}}{x}-2x^3)M_{\dot{x}}+\frac{3}{2x}M=0$ (vide Eq.(\ref{met20})). To obtain particular solutions of this equation we assume the following ansatz for $M$, namely $M=\frac{1}{a(t,x)+b(t,x)\dot{x}+c(t,x)\dot{x}^2}$, where $a, b$ and $c$ are arbitrary functions of their arguments.
Substituting $M$ and its derivatives in the determining equation for $M$ and rearranging the resultant expression we obtain a polynomial equation in $\dot{x}$. Equating the various powers of $\dot{x}$ to zero, we obtain a set of differential equations for the unknowns $a, b$ and $c$. Solving them we can find the explicit forms of $a, b$ and $c$. Substituting them back in the expression for $M$ we can get
\begin{eqnarray}
M_1&=&\frac{2}{\dot{x}^2+4x^4},~~M_2=\frac{1}{2x\dot{x}+t(\dot{x}^2+4x^4)},\label{jlm2}\\
M_3&=&\frac{1}{2tx\dot{x}+2x^2+t^2(\frac{\dot{x}^2} {2}+2x^4)},~~M_4=\frac{1} {x^3}.\label{jlm4}
\end{eqnarray}
The ratio between the JLMs $M_1$ and $M_4$ gives an integral $(I_1)$ and the ratio between the multipliers $M_2$ and $M_4$ yields another integral $(I_2)$ of the form
\begin{eqnarray}
I_1&=&\frac{\dot{x}^2}{x^3}+4x,\quad I_2=\frac{4 x \dot{x}+2t(4x^4+\dot{x}^2)} {x^3}.\label{inte2}
\end{eqnarray}
One can straightforwardly check that $\frac{dI_i} {dt}=0,~i=1,2$, and the integrals $I_1$ and $I_2$ are functionally independent. From these two integrals we can derive the general solution of (\ref{exam}) in the form
\begin{equation}
x=\frac{16I_1}{64+I_2^2-4I_1I_2t+4I_1^2t^2}.\label{soln1}
\end{equation}
We note here that all other ratios of $M_i's,~i=1,2,3,4,$ provide only functionally dependent integrals, that is the resultant integrals turn out to be either a function of $I_1$ or $I_2$ or their combinations, and so no further information is obtained.
  
\subsection{Darboux polynomials~\cite{dar1}}
Let us consider a second order ODE (\ref{main}) which admits a first integral of the form $I=\frac{F(t,x,\dot{x})}{G(t,x,\dot{x})}$, where $F$ and $G$ are functions of their arguments. Taking a total derivative and rewriting the resultant expression, we get
\begin{eqnarray}
\frac{dI}{dt}=\frac{d}{dt}\bigg(\frac{F}{G}\bigg)=0\Rightarrow \dot{F}=g(t,x,\dot{x})F \Rightarrow D[F]=g(t,x,\dot{x})F,\label{dpeq}
\end{eqnarray}
where $D$ is the total differential operator and $g(t,x,\dot{x})=\frac{\dot{G}}{G}$ is the cofactor.
The above equation is the determining equation for the Darboux polynomials \cite{dar1}. Solving Eq.(\ref{dpeq}), we can obtain Darboux polynomials $(F)$ and the cofactors $(g)$. Once sufficient number of Darboux polynomials are known the associated first integral can be determined as follows. The ratio of two Darboux polynomials with the same cofactor defines the first integral \cite{dar1}. For example, let $f_1$ and $f_2$ be two Darboux polynomials with the same cofactor, say $g(t,x,\dot{x})$. In this case, we have
\begin{equation}
\frac{df_1} {dt}=gf_1, ~~~\frac{df_2} {dt}=gf_2.
\end{equation}
Let $I$ be the ratio of two Darboux polynomials, that is $ I=\frac{f_1} {f_2}$. From this expression, we find
\begin{eqnarray}
\frac{dI} {dt}=\frac{d}{dt}\bigg(\frac{f_1} {f_2}\bigg)=0&=&\frac{\dot{f_1}} {f_2}-\frac{\dot{f_2}f_1} {f_2^2},\nonumber \\
0&=&\frac{gf_1} {f_2}-\frac{gf_2f_1} {f_2^2}=0.
\end{eqnarray}
Hence in this method also the necessary integrals can be deduced with sufficient number of polynomials. The combinations of Darboux polynomials also define a first integral.
\subsubsection{Example}
To illustrate the Darboux polynomials approach we again consider the same example given above, that is Eq. (\ref{exam}). The Darboux polynomials can be determined by solving the equation $F_t+\dot{x}F_x+(\frac{3}{2}\frac{\dot{x}}{x}-2x^3)F_{\dot{x}}=gF$ (vide Eq.(\ref{dpeq})). To solve this equation we assume an ansatz for $F$ as $F=a(t,x)+b(t,x)\dot{x}+c(t,x)\dot{x}^2$, where $a,b$ and $c$ are arbitrary functions of their arguments. Substituting this form in (\ref{dpeq}) and rearranging the resultant equation one obtains a polynomial equation  in $\dot{x}$. Equating the various powers of $\dot{x}$ to zero we get a set of equations for the unknowns $a, b$ and $c$. Solving them consistently we can find the exact forms of $a, b$ and $c$. Substituting them back in $F$ we obtain Darboux polynomials of (\ref{exam}) in the form
\begin{eqnarray}
F_1&=&\frac{1}{2}(\dot{x}^2+4x^4),~~F_2=2x\dot{x}+t(\dot{x}^2+4x^4),\\
F_3&=&2tx\dot{x}+2x^2+t^2(\frac{\dot{x}^2} {2}+2x^4),~~F_4=x^3.
\end{eqnarray}
Interestingly all the above Darboux polynomials share the same cofactor $(g)$, that is $\frac{3\dot{x}}{x}$. In other words, one can verify that $D[F_i]=\frac{3\dot{x}}{x}F_i,~i=1,2,3,4$. Since the ratio of two Darboux polynomials which have the same cofactor defines a first integral, we can extract the required integrals from the above Darboux polynomials. For example, the ratio of $F_1$ and $F_4$ provides an integral which  exactly coincides with $I_1$ given in (\ref{inte2}) and the ratio between $F_2$ and $F_4$ gives another integral which in turn matches with the second integral given in Eq.(\ref{inte2}). The other ratios provide only functionally dependent integrals. From these two integrals one can get the general solution (\ref{soln1}).

\subsection{Lie point symmetries~\cite{olv}}
 Let us consider a second order ODE of the form (\ref{main}). The invariance of Eq.(\ref{main}) under an one parameter group of Lie point symmetries, corresponding to infinitesimal transformations,
\begin{eqnarray}
&&T=t+\varepsilon \,\xi(t,x),~~~X=x+\varepsilon \,\eta(t,x),\quad \epsilon \ll 1,\label{asm}
\end{eqnarray}
where $\xi(t,x)$ and $\eta(t,x)$ are the infinitesimal point symmetries and $\varepsilon$ is a small parameter, is given by
\begin{eqnarray}
&&\xi \frac{\partial \phi}{\partial t}+\eta \frac{\partial \phi}{\partial x}+(\eta_t+\dot x (\eta_x-\xi_t)-\dot x^2 \xi_x)\frac{\partial \phi}{ \partial \dot x}-(\eta_{tt}+(2\eta_{tx}-\xi_{tt})\dot x+(\eta_{xx}-2\xi_{tx})\dot x^2\nonumber \\&& \hspace{7cm}-\xi_{xx} \dot x^3+(\eta_x-2\xi_t-3\dot x \xi_x)\ddot x) =0.\label{liec}
\end{eqnarray}
Substituting the known expression $\phi$ in (\ref{liec}) and solving the resultant equation we can get the Lie point symmetries associated with the given ODE. The associated vector field is given by $V=\xi \frac{\partial} {\partial t}+\eta\frac{\partial} {\partial x}$. 

One may also introduce a characteristics $Q=\eta-\dot{x}\xi$ and rewrite the invariance condition (\ref{liec}) in terms of a single variable $Q$ as
\begin{equation}
\frac{d^2Q} {dt^2}-\phi_{\dot{x}}\frac{dQ} {dt}-\phi_x Q=0.
\label{met1411}
\end{equation}
Solving this equation one can get $Q$. From $Q$ one can recover the infinitesimals $\xi$ and $\eta$.

\subsubsection{Example}

To demonstrate the Lie point symmetries approach we again consider the same example given in Eq.(\ref{exam}). The invariance of Eq.(\ref{exam}) under the one parameter Lie group of infinitesimal transformations (\ref{asm}) is given by 
\begin{eqnarray}
&&\eta (\frac{3\dot{x}}{2x^2}+6x^2)-(\eta_t+\dot x (\eta_x-\xi_t)-\dot x^2 \xi_x))\bigg(\frac{3}{2x}\bigg)+\eta_{tt}+(2\eta_{tx}-\xi_{tt})\dot x+(\eta_{xx}-2\xi_{tx})\dot x^2\nonumber \\&&\hspace{6.0cm}-\xi_{xx} \dot x^3+(\eta_x-2\xi_t-3\dot x \xi_x)(\frac{3}{2}\frac{\dot{x}}{x}-2x^3)=0.\label{invar-exa}
\end{eqnarray}
Equating the coefficients of various powers of $\dot{x}^m, m=0,1,2,3$, to zero and solving the resultant partial differential equations we can find the explicit forms of $\xi$ and $\eta$ which turn out to be $\xi=b_1+b_2t+b_3t^2$ and $\eta=-(b_2x+2b_3tx)$, where $b_i's,i=1,2,3$, are arbitrary constants. The associated vector fields can be written as ($V=\xi \frac{\partial} {\partial t}+\eta\frac{\partial} {\partial x}$)
\begin{eqnarray}
V_1&=&\frac{\partial}{\partial t},~~V_2=t\frac{\partial}{\partial t}-x \frac{\partial}{\partial x},~~ V_3=t^2 \frac{\partial}{\partial t}-2tx\frac{\partial}{\partial x}.
\end{eqnarray}

\subsection{Adjoint symmetries~\cite{blum}}
In general, for systems of one or more ODEs, an integrating factor is a set of functions, multiplying each of the ODEs, which yields a first integral. If the system is self-adjoint, then its integrating factors are necessarily solutions of its linearized system. Such solutions are the symmetries of the given system of ODEs. If a given system of ODEs is not self-adjoint, then its integrating factors are necessarily solutions of the adjoint system of its linearized system. Such solutions are known as adjoint symmetries of the given system of ODEs \cite{blum}. The adjoint ODE of linearized symmetry condition (\ref{met1411}) can be written as
\begin{equation}
\frac{d^2\Lambda} {dt^2}+\frac{d} {dt}(\phi_{\dot{x}}\Lambda)-\phi_x\Lambda=0.\label{adjk1}
\end{equation}
Evaluating Eq.(\ref{adjk1}) and collecting the powers of $\ddot{x}$, we get
\begin{eqnarray}
\Lambda_{t\dot{x}}+\Lambda_{x\dot{x}}\dot{x}+2\Lambda_{x}+\Lambda \phi_{\dot{x}\dot{x}}+2\phi_{\dot{x}}\Lambda_{\dot{x}}+\phi \Lambda_{\dot{x}\dot{x}}&=&0,\label{adjo1}\\
\Lambda_{tt}+2\Lambda_{tx}\dot{x}+\Lambda_{xx}\dot{x}^2+\Lambda \phi_{t\dot{x}}+\Lambda \phi_{x\dot{x}}\dot{x}+\phi_{\dot{x}}\Lambda_{t} +\phi_{\dot{x}}\Lambda_{x}\dot{x}\nonumber \\-\Lambda \phi_{x}-\Lambda_{x}\phi+\phi \Lambda_{t\dot{x}}+\phi \Lambda_{x\dot{x}}\dot{x} +\Lambda_{\dot{x}}\phi_t+\Lambda_{\dot{x}}\phi_{x}\dot{x}&=&0.\label{adjo2}
\end{eqnarray}
Solutions of the Eq.(\ref{adjo2}) are called adjoint symmetries. If these solutions also satisfy Eq.(\ref{adjo1}) then they become integrating factors. Once we know the integrating factors, we can find the first integrals. Multiplying the given equation by these integrating factors, and rewriting the resultant equation as a perfect differentiable function, we can identify the first integrals, that is
\begin{equation}
\Lambda_i(t,x,\dot{x})(\ddot{x}-\phi(t,x,\dot{x}))=\frac{d} {dt}I_i,~i=1,2,...m.\label{adjint}
\end{equation}
\subsubsection{Example}
If the symmetry determining equation (\ref{met1411}) is the same as its adjoint determining equation (\ref{adjk1}), then the adjoint-symmetries are symmetries. In this case, the equation is called self-adjoint equation. If the given equation is not self adjoint, then the solutions of Eq.(\ref{adjk1}) are called adjoint symmetries. Since the example (\ref{exam}) is not self-adjoint we have to find the adjoint symmetries. So we have to solve the following differential equation $\frac{d^2\Lambda} {dt^2}+\frac{d} {dt}(\frac{3} {2x}\Lambda)+(\frac{3\dot{x}} {2x^2}+6x^2)\Lambda=0$. To achieve this task we assume an ansatz for $\Lambda$ as $\Lambda=\frac{a_1(t,x)+a_2(t,x)\dot{x}}{b_1(t,x)+b_2(t,x)\dot{x}+b_3(t,x)\dot{x}^2}$, where $a_i's$ and $b_j's,i=1,2,j=1,2,3,$ are all to be determined. Substituting this ansatz in (\ref{adjk1}) with $\phi=\frac{3\dot{x}}{2x}-2x^3$ and solving the resultant equation we can get the following particular solutions, that is
\begin{eqnarray}
\Lambda_1=\frac{-2\dot{x}}{x^3},~~\Lambda_2=\frac{-(x+t\dot{x})}{4tx^4+2 x \dot{x}+t\dot{x}^2}.
\end{eqnarray}
We find that these two expressions also satisfy Eq.(\ref{adjo1}) and so they become the integrating factors of (\ref{exam}) as well. Multiplying the given Eq.(\ref{exam}) by these integrating factors separately and rewriting them as a perfect derivative we can find the same integrals $I_1$ and $I_2$ given in (\ref{inte2}). 

\subsection{$\lambda$-symmetries approach~\cite{mur}}
All the nonlinear ODEs do not necessarily admit Lie point symmetries.  Under such a circumstance one may look for  generalized symmetries (other than Lie point symmetries) associated with the given equation.  One such generalized symmetry is the $\lambda$-symmetry. Let $\tilde{V}=\xi(t,x)\frac{\partial}{\partial t}+\eta(t,x)\frac{\partial}{\partial x}$ be a $\lambda$-symmetry of the given ODE for some function $\lambda=\lambda(t,x,\dot{x})$. The invariance of the given ODE under $\lambda$-symmetry vector field is given by $V^{[\lambda,(2)]}(\ddot{x}-\phi(t,x,\dot{x}))=0$, where $V^{[\lambda,(2)]}$ is given by $\xi \frac{\partial} {\partial t}+\eta \frac{\partial} {\partial x}+\eta^{[\lambda,(1)]}\frac{\partial} {\partial \dot{x}}+\eta^{[\lambda,(2)]}\frac{\partial} {\partial \ddot{x}}$. Here $\eta^{[\lambda,(1)]}$ and $\eta^{[\lambda,(2)]}$ are first and second $\lambda$- prolongations. They are given by
\begin{eqnarray}
\eta^{[\lambda,(1)]}&=&(D_t+\lambda)\eta^{[\lambda,(0)]}(t,x)-(D_t+\lambda)(\xi(t,x))\dot{x},\label{etlam1} \nonumber\\
\eta^{[\lambda,(2)]}&=&(D_t+\lambda)\eta^{[\lambda,(1)]}(t,x,\dot{x})-(D_t+\lambda)(\xi(t,x))\ddot{x}.\label{etlam2}
\end{eqnarray}
Expanding the invariance condition, we find
\begin{equation}
\xi\phi_t+\eta\phi_x+\eta^{[\lambda,(1)]}\phi_{\dot{x}}-\eta^{[\lambda,(2)]}=0.
\label{beq1}
\end{equation}
If we put $\lambda=0$, we get the Lie prolongation formula. Solving the invariance condition (\ref{beq1}) we can obtain the explicit forms of $\xi, \eta$ and $\lambda$. Once the functions $\xi, \eta$ and $\lambda$ are known the integrals can be evaluated using the following procedure \cite{mur}. 
\begin{enumerate}
\item
Find a first integral $w(t,x,\dot x)$ of $v^{[\lambda,(1)]}$, that is a particular solution of the equation
\begin{equation}
w_x+\lambda w_{\dot{x}}=0,
\label{weq1}
\end{equation}
where subscripts denote partial derivative with respect to that variable and $v^{[\lambda,(1)]}$ is the first order $\lambda$-prolongation of the vector field $v$.
\item
Evaluate $A(w)$ and express $A(w)$ in terms of $(t,w)$ as $A(w)=F(t,w)$.
\item
Find a first integral G of $\partial_t+F(t,w)\partial_w$.
\item
Evaluate $I(t,x,\dot{x})=G(t,w(t,x,\dot{x}))$.
\end{enumerate}
If the given ODE admits Lie point symmetries, then the $\lambda$-symmetries can be derived without solving the $\lambda$-prolongation condition. In this case the $\lambda$-symmetries can be uniquely deduced from
\begin{equation}
\lambda=\frac{D[Q]} {Q},\label{lanb}
\end{equation}
where $D$ is the total differential operator and $Q=\eta-\xi\dot{x}$. The associated vector field is given by $V=\frac{\partial} {\partial x}$.
\subsubsection{Example}
To illustrate the method we again consider the same example, namely Eq.(\ref{exam}). In the first step we determine the $\lambda$-symmetries of this equation. Since we already know the system admits Lie point symmetries, one can capture the $\lambda$-symmetries straightforwardly by recalling the relation (\ref{lanb}). The resultant expressions read
\begin{eqnarray}
\lambda_1&=&\frac{3\dot{x}}{2x}-\frac{2x^3}{\dot{x}}, ~~\lambda_2=\frac{4 x \dot{x}+3 t \dot{x}^2-4 t x^4} {2 x^2+ t x \dot{x}},~~
\lambda_3=\frac{1}{t}+\frac{3\dot{x}}{2x}-\frac{2tx^3}{2x+t\dot{x}}.\label{lam3}
\end{eqnarray}
With $\xi=0$ and $\eta=1$, one can easily verify that the invariance condition (\ref{beq1}) is satisfied for all the above three forms $\lambda_1, \lambda_2$ and $\lambda_3$. The associated vector field is given by $V=\frac{\partial} {\partial x}$. In the second step we have to derive the integral associated with each $\lambda$-symmetry which can be done through the algorithm given above. For illustration purpose we consider only one case. For the remaining two cases one can adopt the same procedure and derive the associated integrals.

Substituting $ \lambda_1 = \frac{3\dot{x}}{2x}-\frac{2x^3}{\dot{x}}$ in (\ref{weq1}) one gets
\begin{equation}
w_x+\left(\frac{3\dot{x}}{2x}-\frac{2x^3}{\dot{x}}\right)w_{\dot{x}}=0.
\label{lam1eq1}
\end{equation}
To obtain a particular solution of (\ref{lam1eq1}) we solve the associated characteristic equation, namely
\begin{equation}
\frac{dx}{1} = \frac{d\dot x}{\frac{3\dot{x}}{2x}-\frac{2x^3}{\dot{x}}}.
\label{lam1eq2}
\end{equation}
Integrating (\ref{lam1eq2}) we find a first integral $w(t,x,\dot{x})$ which is of the form
\begin{equation}
w(t,x,\dot{x}) = \frac{\dot{x}^2}{x^3}+4x.
\label{lam1eq3}
\end{equation}
In the second step one has to evaluate $A(w)$, that is
\begin{equation}
A(w) = \frac{\partial w} {\partial t}+\dot{x} \frac{\partial w} {\partial x}+(\frac{3}{2}\frac{\dot{x}}{x}-2x^3)\frac{\partial w} {\partial \dot{x}}=0\Rightarrow A(w) =F=0.
\label{lam1eq4}
\end{equation}
Since $F=0$, the function $w$ becomes the integral $I_1$. In other words we have $I_1=\frac{\dot{x}^2}{x^3}+4x$. The integral $I_1$ exactly matches with the one given in Eq.(\ref{inte2}).
In the same way, one can get the other two integrals from the $\lambda_2$ and $\lambda_3$, respectively of the form
\begin{eqnarray}
I_2=\frac{4tx^4+2x\dot{x}+t\dot{x}^2}{x^3}, ~~I_3=\frac{4x^2+4t^2x^4+4tx\dot{x}+t^2\dot{x}^2} {x^3}.
\label{i3}
\end{eqnarray}
Here also one can check that the total derivatives vanish, that is $\frac{dI_i} {dt}=0,~~i=1,2,3$. In the above $I_1$ and $I_2$ are the two independent integrals and $I_3=\frac{I_2^2+64} {4I_1}$. The general solution (\ref{soln1}) can be found from the integrals $I_1$ and $I_2$ given above. 

\subsection{Prelle-Singer method~\cite{chan,duar,prelle}}

 Let us assume that the ODE (\ref{main})
admits a first integral $I(t,x,\dot{x})=C,$ with $C$ constant on the
solutions, so that the total differential gives
\begin{eqnarray}
dI={I_t}{dt}+{I_{x}}{dx}+{I_{\dot{x}}{d\dot{x}}}=0,
\label{met3}
\end{eqnarray}
where the subscript denotes partial differentiation with respect to that variable.  Rewriting equation~(\ref{main}) in the form
$\phi dt-d\dot{x}=0$ and adding a null term
$S(t,x,\dot{x})\dot{x}$ $ dt - S(t,x,\dot{x})dx$ to the latter, we obtain that on
the solutions the 1-form
\begin{eqnarray}
\bigg(\frac{P}{Q}+S\dot{x}\bigg) dt-Sdx-d\dot{x} = 0.
\label{met6}
\end{eqnarray}	
Hence, on the solutions, the 1-forms (\ref{met3}) and
(\ref{met6}) must be proportional.  Multiplying (\ref{met6}) by the
factor $ R(t,x,\dot{x})$ which acts as the integrating factor
for (\ref{met6}), we have on the solutions that
\begin{eqnarray}
dI=R(\phi+S\dot{x})dt-RSdx-Rd\dot{x}=0.
\label{met7}
\end{eqnarray}
Comparing equations (\ref{met3})
with (\ref{met7}) we end up with the three relations which relates the integral($I$), integrating factor($R$) and the null term($S$),
\begin{eqnarray}
I_{t}=R(\phi+\dot{x}S),~I_{x} =-RS,~I_{\dot{x}} = -R.
\label{met8}
\end{eqnarray}
The compatibility conditions between these relations that is $I_{tx}=I_{xt}$, $I_{t\dot{x}}=I_{{\dot{x}}t}$,
$I_{x{\dot{x}}}=I_{{\dot{x}}x}$ lead us to the following determining equations for the unknowns $R$ and $S$, namely
\begin{eqnarray}
D[S]=&-\phi_x+S\phi_{\dot{x}}+S^2,\label{met9}\\
D[R]=&-R(S+\phi_{\dot{x}}),\label{met10}\\
R_x=&R_{\dot{x}}S+RS_{\dot{x}},\label{met11}
\end{eqnarray}
where $D=\frac{\partial}{\partial{t}}+
\dot{x}\frac{\partial}{\partial{x}}+\phi\frac{\partial}
{\partial{\dot{x}}}$.
Equations~(\ref{met9})-(\ref{met11}) can be solved in principle in the following way.
From (\ref{met9}) we can find $S$. Once $S$ is known then
(\ref{met10}) becomes the determining equation for the function $R$.
Solving the latter, one can get an explicit form for $R$.
Now the functions $R$ and $S$ have to satisfy an extra constraint, that is
equation~(\ref{met11}).  Once a compatible solution satisfying all the three
equations have been found then the functions $R$ and $S$ fix the integral of motion
$I(t,x,\dot{x})$ by the relation
\begin{eqnarray}
I(t,x,\dot{x})= \int R(\phi+\dot{x}S)dt
 -\int \left( RS+\frac{d}{dx}\int R(\phi+\dot{x}S)dt \right) dx \nonumber\\
 -\int \left\{R+\frac{d}{d\dot{x}} \left[\int R (\phi+\dot{x}S)dt-
\int \left(RS+\frac{d}{dx}\int R(\phi+\dot{x}S)dt\right)dx\right]
\right\}d\dot{x}. \label{met13}
\end{eqnarray}
Equation~(\ref{met13}) can be derived straightforwardly by
integrating the three relations which relates the integral($I$), integrating factor($R$) and the null term($S$).
Note that for every independent set $(S,R)$, equation~(\ref{met13}) defines an integral. Once we find  sufficient number of independent integrals (for example two independent integrals for a second order nonlinear ODE) we can write the general solution. In the following, we illustrate the method with the example (\ref{exam}).

\subsubsection{Example}
To obtain the integrals and the general solution of (\ref{exam}), in the PS method, we have to find null forms and their associated integrating factors. To obtain the null form we have to solve the following equation, namely
\begin{eqnarray}
S_t+\dot{x}S_x+(\frac{3}{2}\frac{\dot{x}}{x}-2x^3)S_{\dot{x}}=\frac{3} {2}\frac{\dot{x}} {x^2}+S\frac{3} {2x}+S^2.
\end{eqnarray}
To solve this equation we assume an ansatz for $S$ in the form $S=\frac{a_1(t,x)+a_2(t,x)\dot{x}+a_3(t,x)\dot{x}^2}{b_1(t,x)+b_2(t,x)\dot{x}+b_3(t,x)\dot{x}^2}$, where $a_i's,$ and $b_i's,i=1,2,3$ are functions of their arguments. Substituting this ansatz in the $S$ determining equation and rearranging the resultant equations we get a polynomial equation in $\dot{x}$. Equating the coefficients of various powers of $\dot{x}$, to zero we get a set of partial differential equations for the unknown functions $a_i's$ and $b_i's$. Solving them we obtain their explicit forms. In our case we find the following two particular solutions, namely
\begin{eqnarray}
S_1&=&\frac{2x^3}{\dot{x}}-\frac{3\dot{x}}{2x},~~S_2=\frac{4 t x^4-4 x \dot{x}-3 t \dot{x}^2} {2 x^2+ t x \dot{x}}.\label{s2}
\end{eqnarray}
Once we know the null forms, substituting them separately in (\ref{met10}) and solving the resultant equation we can get the integrating factors. In both the cases we assume an ansatz for $R$ in the form
\begin{equation}
R=\frac{a_1(t,x)+a_2(t,x)\dot{x}}{b_1(t,x)+b_2(t,x)\dot{x}+b_3(t,x)\dot{x}^2},\label{rearrv}
\end{equation}
where $a_i's$ and $b_j's$, $i=1,2,~j=1,2,3$, are arbitrary functions of their arguments. With this ansatz we solve the Eq. (\ref{met10}) and obtain the integrating factors in the following form
\begin{eqnarray}
R_1&=&\frac{-2\dot{x}}{x^3},~~R_2=\frac{-(x+t\dot{x})}{4tx^4+2 x \dot{x}+t\dot{x}^2}.\label{rr2}
\end{eqnarray}
The expressions $(S_i,~R_i),~i=1,2$, satisfy the third equation (\ref{met11}) also. With the help of these null forms and integrating factors, we can find the integrals by substituting them in (\ref{met13}) and evaluating the integrals. Doing so we find the same integrals as given in Eq.(\ref{inte2}). From these two integrals one can obtain the general solution (\ref{soln1}).

\section{Interconnections}
In the previous section we illustrated certain methods which are used to derive integrating factors, symmetries and integrals of a given dynamical system. In this section we demonstrate from null forms and integrating factors of PS procedure one can recover Lie point symmetries, $\lambda$-symmetries, adjoint symmetries, Darboux polynomials and Jacobi multipliers.
To demonstrate that the null form $S$ and the integrating factor $R$ are intimately related with other measures of integrability we identify the following relations. 
By introducing a transformation
\begin{equation}
S=-D[X]/X\label{sx},
\end{equation} Eq.(\ref{met9}) becomes a linear equation in the new variable $X$, that is
\begin{equation}
D^2[X]  = \phi_{\dot{x}} D[X]+\phi_x X,
\label{met14}
\end{equation}
where $D$ is the total differential operator. With another change of variable
\begin{equation}
R=X/F,\label{rxf}
\end{equation}
where $F(t,x,\dot{x})$ is a function to be determined, we can rewrite Eq.(\ref{met10}) in a compact form in the new variable $F$ as
\begin{equation}
D[F] = \phi_{\dot{x}}F.
\label{met15}
\end{equation}

In the following, we demonstrate that the functions $X$ and $F$ are intimately related to Lie symmetries, $\lambda$-symmetries, Darboux polynomials, Jacobi last multiplier and adjoint symmetries. By doing this we also capture the known connections. For more details, see our recent work \cite{subha}.

\subsection{Connection between $\lambda$- symmetries and null forms}
If we replace $S=-Y$ in equation (\ref{met9}) we get
\begin{equation}
D[Y]  = \phi_x+Y\phi_{\dot{x}}-Y^2,\label{met17}
\end{equation}
which is nothing but the determining equation for the $\lambda$-symmetries for a second order ODE \cite{mur} which in turn establishes the connection between $\lambda$-symmetries and null forms. It has also been shown \cite{mur} that once Lie point symmetries are known, the $\lambda$-symmetries can be constructed through the relation $Y=D[Q]/Q$ which is also confirmed here.
\subsection{Connection between JLM and Lie point symmetries}
The Jacobi last multiplier can also be calculated from the Lie point symmetries by evaluating the determinant \cite{jaco1,jaco2}
\begin{equation}
 \Delta = \left| \begin{array}{ccc}
1 & \dot{x} & \ddot{x} \\
\xi_1 & \eta_1 & \eta_{1}^{(1)} \\
\xi_2 & \eta_2 & \eta_{2}^{(1)} \end{array} \right|,\label{det1}
\end{equation}
where $(\xi_1, \eta_1)$ and $(\xi_2,\eta_2)$ are the infinitesimal Lie point symmetries and $\eta_{1}^{(1)}$ and $\eta_{2}^{(1)}$ are their corresponding first prolongations. Inverting the determinant we get Jacobi last multiplier $M=\big(\frac{1} {\Delta}\big)$. The determinant establishes the connection between multiplier and Lie point symmetries \cite{jaco1,jaco2}.

\subsection{Connection between Lie point symmetries and null forms}

We have seen that the invariance condition $v^{(2)}[\ddot{x}-\phi(t,x,\dot{x})]=0$ determines the infinitesimal symmetries $\xi$ and $\eta$ explicitly \cite{olv}. The invariance condition in terms of the evolutionary vector field $Q$ is given by
\begin{equation}
D^2[Q]  = \phi_{\dot{x}} D[Q]+\phi_x Q.
\label{met16}
\end{equation}
Comparing Eqs.(\ref{met14}) and (\ref{met16}) we find that
\begin{equation}
X=Q.\label{xq}
\end{equation}
In other words the $S$-determining equation (\ref{met14}) now becomes exactly the determining equation for the Lie point symmetries $\xi$ and $\eta$ with $Q=\eta-\dot{x}\xi$. Since $S=-\frac{D[X]} {X}$, the null form $S$ can also be determined once $\xi$ and $\eta$ are known. This establishes the connection between the null forms $S$ with Lie point symmetries ($\xi$ and $\eta$).

\subsection{Darboux polynomials and integrating factors}
 Let us consider the function $G=\prod f_i^{n_i}$, where $f_i'$s are Darboux polynomials and $n_i'$s are rational numbers. If we can identify a sufficient number of Darboux polynomials (irreducible polynomials) $f_i'$s, satisfying the relations $D[f_i]/f_i=\alpha_i$, where $\alpha_i$'s are the co-factors, then
\begin{equation}
D[G]/G=\sum_in_i\frac{D[f_i]}{f_i}=n_i\alpha_i.\label{laseq}
\end{equation}
Now we compare Eq.(\ref{laseq}) with Eq.(\ref{met15}). If we choose $G=F$ then equation (\ref{met15}) becomes
\begin{equation}
\sum_in_i\frac{D[f_i]}{f_i}=\phi_{\dot{x}}.\label{met22}
\end{equation}
In other words the Darboux polynomials constitute the solution of Eq.(\ref{met15}). Since $X$ is already known, once Darboux polynomials are known the integrating factors can be fixed (vide Eq.(\ref{rxf})). 

\subsection{JLM and Darboux polynomials/Integrating factors}
Comparing the equation (\ref{met15}) with (\ref{met20}) we find that
\begin{equation}
F=M^{-1}.\label{hjfdh}
\end{equation}
 Thus from the knowledge of the multiplier $M=\big(\frac{1} {\Delta}\big)$ we can also fix the explicit form of $F$ which appears in the denominator of integrating factor $R$ (vide Eq.(\ref{rxf})) as $F=\Delta$ or vice versa.

\subsection{Adjoint symmetries and integrating factors}
Let us rewrite the coupled equations (\ref{met9}) and (\ref{met10}) into an equation for the single function $R$. Then the resultant equation turns out to be of the form 
\begin{equation}
D^2[R]+D[\phi_{\dot{x}}R]-\phi_{x}R=0.\label{met24}
\end{equation}

Comparing the above two equations (\ref{adjk1}) and (\ref{met24}) one can conclude that the integrating factor $R$ is nothing but the adjoint symmetry $\Lambda$, that is
\begin{equation}
R=\Lambda. \label{met25}
\end{equation}
Thus the integrating factor turns out to be the adjoint symmetry of the given second order ODE.

\subsection{Example}
We illustrate the interconnections by considering the same example (\ref{exam}). In the above, we have shown that the interconnections between the several analytical methods to solve second order differential equations. Since all the quantities are interrelated we start with the Lie point symmetries. From the Lie point symmetries, $V_1$ and $V_2$, we can determine a multiplier by evaluating the determinant
 \begin{eqnarray}
\Delta_{12} &=&  \left| \begin{array}{ccc}
1 & \dot{x} & \ddot{x} \\
\xi_1 & \eta_1 & \eta_{1}^{(1)} \\
\xi_2 & \eta_2 & \eta_{2}^{(1)} \end{array} \right|=\begin{vmatrix}
1 & \dot{x} & \ddot{x} \\
1 & 0 & 0 \\
t & -x & -2\dot{x} \end{vmatrix}=\frac{1}{2}(\dot{x}^2+4x^4).\label{delta}
\end{eqnarray}
The reciprocal of this determinant $(\frac{1} {\Delta_{12}})$ gives a last multiplier which coincides exactly with $M_1$ (vide Eq.(\ref{jlm2})). From the Lie point symmetries $V_1$ and $V_3$, we can get $\Delta_{13}$ and from $V_2$ and $V_3$ we can get $\Delta_{23}$. The expressions $\frac{1} {\Delta_{13}}$ and $\frac{1} {\Delta_{23}}$ defines the multipliers $M_2$ and $M_3$ respectively which in turn exactly matches with expressions given in (\ref{jlm2}) and (\ref{jlm4}). From the null forms (\ref{s2}), we can fix the $\lambda$-symmetries as $S=-Y$. The adjoint symmetries are nothing but the integrating factors has already been observed through the expressions (\ref{met25}) and (\ref{rr2}). The reciprocals of multipliers provide the Darboux polynomials which can be straightforwardly verified in this example.
\section{Conclusion}
In this paper, we have considered six different analytical methods which are widely used in the contemporary literature. In particular, we have considered (i) Jacobi last multiplier, (ii) Darboux polynomials, (iii) Lie point symmetries, (iv) adjoint symmetries, (v) $\lambda$-symmetries and (vi) Prelle-Singer method. To begin with we have recalled each one of the methods separately with an example. The same example is considered for all the methods in order to illustrate the merits and demerits of each method compared to other methods. These methods essentially seek either one or more of the following quantities, namely (i) symmetries, (ii) integrating factors, (iii) multipliers, (iv) Darboux polynomials and (v) integrals. We have investigated the interlinks between these methods in detail. After performing a thorough analysis we have shown that starting from the null form and the integrating factor in the Prelle-Singer procedure one can obtain $\lambda$-symmetries, Lie point symmetries, Darboux polynomials and adjoint symmetries. We have explicitly shown the interconnections between these factors. The same example has been considered to illustrate the interlinks also.

\section{Acknowledgement}
MS and ML wish to thank the organizer Dr. Swapan Kumar Ghosh of the CNPA-13 meeting for his excellent hospitality. RMS acknowledges the University Grants Commission (UGC-RFSMS), Government of India, for providing a Research Fellowship and the work of MS forms part of a research project sponsored by Department of Science and Technology, Government of India. The work of ML is supported by a Department of Science and Technology (DST), Government of India, IRHPA research project. ML is also supported by a DAE Raja Ramanna Fellowship and a DST Ramanna Fellowship programme.

\end{document}